\documentclass[12pt]{article}
\usepackage{graphicx}
\usepackage{epsf}
\newcommand{\rnc}{\renewcommand}

\makeatletter \@addtoreset{equation}{section}
\rnc{\theequation}{\thesection.\arabic{equation}} \makeatother

\def\sqr#1#2{{\vcenter{\vbox{\hrule height.#2pt
\hbox{\vrule width.#2pt height#1pt \kern#1pt \vrule
width.#2pt}\hrule height.#2pt}}}}
\def\square{\mathchoice\sqr45\sqr45\sqr{2.1}3\sqr{1.5}3}

\def\a{\alpha}
\def\b{\beta}

\def\d{\delta}\def\D{\Delta}
\def\e{\epsilon}

\def\f{\phi}\def\F{\Phi}
\def\bphi{{\bar\phi}}
\def\h{\eta}
\def\k{\kappa}
\def\l{\lambda}\def\L{\Lambda}
\def\m{\mu}
\def\n{\nu}
\def\r{\rho}
\def\s{\sigma}\def\S{\Sigma}
\def\t{\tau}

\def\X{\Xeta}
\def\be{\begin{equation}}\def\ee{\end{equation}}
\def\bea{\begin{eqnarray}}\def\eea{\end{eqnarray}}
\def\ba{\begin{array}}\def\ea{\end{array}}
\def\x{\xi}

\let\la=\label
\let\bl=\bigl \let\br=\bigr
\let\Br=\Bigr \let\Bl=\Bigl
\let\bm=\bibitem

\def\nn{\nonumber}
\def\bd{\begin{document}}
\def\ed{\end{document}}
\def\ft#1#2{{\textstyle{{\scriptstyle #1}\over {\scriptstyle #2}}}}
\def\fft#1#2{{#1 \over #2}}
\newcommand{\eq}[1]{(\ref{#1})}
\def\eqs#1#2{(\ref{#1}-\ref{#2})}
\def\det{{\rm det\,}}
\def\tr{{\rm tr}}
\def\Tr{{\rm Tr}}
\newcommand{\ho}[1]{$\, ^{#1}$}
\newcommand{\hoch}[1]{$\, ^{#1}$}
\def\fdf{\phi^\dagger\phi}
\def\ffd{\phi\phi^\dagger}
\def\qq{\quad\quad}
\newcommand{\w}[1]{\\[0.#1cm]}
\def\lra{\leftrightarrow}

\def\cramp{\medmuskip = 2mu plus 1mu minus 2mu}
\def\cramper{\medmuskip = 2mu plus 1mu minus 2mu}
\def\crampest{\medmuskip = 1mu plus 1mu minus 1mu}
\def\uncramp{\medmuskip = 4mu plus 2mu minus 4mu}
\def\ben{\begin{equation}}
\def\een{\end{equation}}
\def\half{{\textstyle{1\over2}}}
\let\a=\alpha \let\b=\beta \let\g=\gamma \let\d=\delta \let\e=\epsilon
\let\z=\zeta \let\h=\eta \let\q=\theta \let\i=\iota \let\k=\kappa
\let\l=\lambda \let\m=\mu \let\n=\nu \let\x=\xi
\let\r=\rho
\let\s=\sigma \let\t=\tau \let\f=\phi  \let\y=\psi
\let\D=\Delta \let\Q=\Theta \let\L=\Lambda
\let\X=\Xi \let\P=\Pi \let\S=\Sigma \let\U=\Upsilon \let\F=\Phi \let\Y=\Psi
\let\W=\Omega
\let\la=\label \let\ci=\cite \let\re=\ref
\let\se=\section \let\sse=\subsection \let\ssse=\subsubsection
\def\nn{\nonumber} \def\bd{\begin{document}} \def\ed{\end{document}}
\def\ds{\documentstyle} \let\fr=\frac \let\bl=\bigl \let\br=\bigr
\let\Br=\Bigr \let\Bl=\Bigl
\let\bm=\bibitem
\let\na=\nabla
\let\pa=\partial \let\ov=\overline
\def\ba{\begin{array}}
\def\ea{\end{array}}
\def\ft#1#2{{\textstyle{{\scriptstyle #1}\over {\scriptstyle #2}}}}
\def\fft#1#2{{#1 \over #2}}
\def\del{\partial}
\def\vp{\varphi}
\def\sst#1{{\scriptscriptstyle #1}}
\def\oneone{\rlap 1\mkern4mu{\rm l}}
\def\td{\tilde}
\def\ie{\rm i.e.\ }
\def\dalemb#1#2{{\vbox{\hrule height .#2pt
        \hbox{\vrule width.#2pt height#1pt \kern#1pt
                \vrule width.#2pt}
        \hrule height.#2pt}}}
\def\square{\mathord{\dalemb{6.8}{7}\hbox{\hskip1pt}}}
\newcommand{\ap}{\alpha^\prime}
\newcommand{\bp}{\tilde \beta^\prime}
\def\0{{\sst{(0)}}}
\def\1{{\sst{(1)}}}
\def\2{{\sst{(2)}}}
\def\3{{\sst{(3)}}}
\def\4{{\sst{(4)}}}
\def\5{{\sst{(5)}}}
\def\6{{\sst{(6)}}}
\def\7{{\sst{(7)}}}
\def\8{{\sst{(8)}}}
\def\n{{\sst{(n)}}}
\def\cA{{{\cal A}}}
\def\cF{{{\cal F}}}
\def\tV{\widetilde V}
\def\tW{\widetilde W}
\def\tH{\widetilde H}
\def\tE{\widetilde E}
\def\tF{\widetilde F}
\def\tA{\widetilde A}
\def\im{{{\rm i}}}
\def\tY{{{\wtd Y}}}
\def\ep{{\epsilon}}
\def\vep{{\varepsilon}}
\def\R{\rlap{\rm I}\mkern3mu{\rm R}}
\def\bD{{{\bar D}}}

\def\R{\rlap{\rm I}\mkern3mu{\rm R}}
\def\bD{{{\bar D}}}
\def\R{{{\Bbb R}}}
\def\CN{{{\Bbb C}}}
\def\H{{{\Bbb H}}}
\def\CP{{{\Bbb C}{\Bbb P}}}
\def\RP{{{\Bbb R}{\Bbb P}}}
\def\Z{{{\Bbb Z}}}
\def\bA{{{\Bbb A}}}
\def\bB{{{\Bbb B}}}
\def\bC{{{\Bbb C}}}
\def\bR{{{\Bbb R}}}
\def\bD{{{\Bbb D}}}
\def\bE{{{\Bbb E}}}
\def\bZ{{{\Bbb Z}}}
\def\Re{{{\frak{Re}}}}
\def\Im{{{\frak{Im}}}}
\def\cosec{{\,\hbox{cosec}\,}}
\def\Gm{{\Gamma_{\!\! -}}}
\def\Gp{{\Gamma_{\!\! +}}}
\def\stan{{standard }}
\def\nonstan{{supernumerary }}

\def\cosech{{\hbox{cosech}}}

\def\etcyc{{\hbox{and cyclic}}}
\def\btheta{{\bar\theta}}

\def\bphi{\overline{\phi}}

\newcommand{\tamphys}{\it\small Department of Physics,
Brown University, Providence, RI 02912,  USA}

\newcommand{\ictp}{\it\small International Center for Theoretical
Physics, 34100 Trieste, Italy}

\begin{document}

\begin{flushright}
IC/2004/17 \\
BROWN-HET-1399 \\
MCGILL-04-09 \\
\end{flushright}

\begin{center}
\bigskip
{\Large \bf Inflation in 6D Gauged Supergravity} \\
\bigskip
\bigskip
{\large 
Robert H. Brandenberger
$^{a,b,c,}$\footnote{rhb@het.brown.edu} and 
S. Randjbar-Daemi $^{d,}$\footnote{seif@ictp.trieste.it}}
\end{center}

\begin{flushleft}
~~~~~~~~~~~$a$: {\it Department of Physics, Brown University,
Providence, RI 02912, USA}\\
~~~~~~~~~~~$b$: {\it Perimeter Institute for Theoretical Physics,}\\
~~~~~~~~~~~~~~~{\it Waterloo ON, N2J 2W9, CANADA}\\
~~~~~~~~~~~$c$: {\it Physics Department, McGill University,}\\
~~~~~~~~~~~~~~~{\it 3600 rue University , Montr\'eal QC, H3A 2T8,
CANADA}\\
~~~~~~~~~~~$d$: {\it International Center for Theoretical Physics,
34100 Trieste, ITALY}
\end{flushleft}

\begin{center}
\bigskip (\today)
\vskip0.5cm {\Large Abstract\\} \vskip3truemm
\parbox[t]{\textwidth} {In this note we demonstrate that chaotic
inflation can naturally be realized in the context of an anomaly
free minimal gauged  supergravity in D=6 which has recently been
the focus of some attention. This particular model has a unique
maximally symmetric ground state solution, $R^{3,1}\times S^2$
which leaves half of the six-dimensional supersymmetries unbroken.
In this model, the inflaton field $\phi$ originates from the
complex scalar fields in the D=6 scalar hyper-multiplet. The mass
and the self couplings of the scalar field are dictated by the D=6
Lagrangian. The scalar potential has an absolute minimum at
$\phi=0$ with no undetermined moduli fields. Imposing a mild bound
on the radius of $S^2$ enables us to obtain chaotic inflation. The
low energy equations of motion are shown to be consistent for the
range of scalar field values relevant for inflation.}
\end{center}

\thispagestyle{empty}
\newpage
\setcounter{page}{1}
% Decrease texheight (for preprint numbers) again
%\textheight 23.0 true cm
\baselineskip=14pt

%%%%%%%%%%%%%%%%%%%%%%%%%%%%%%%%%%%%%%%%%%%%%%%%%%%%%%%%%%%%%%%%%%%%%%%%%

\section{Introduction}

%%%%%%%%%%%%%%%%%%%%%%%%%%%%%%%%%%%%%%%%%%%%%%%%%%%%%%%%%%%%%%%%%%%%%%%%%

Cosmological inflation is the most widely accepted solution
\cite{Guth} to the famous cosmological problems of standard big
bang cosmology, and it provides a compelling mechanism for the
origin of cosmological fluctuations (see e.g. \cite{MFB} for a
comprehensive review). Despite its successes, we still lack a
compelling model in which inflation can be realized without
unnatural fine tunings. The origin of the scalar field which
triggers inflation and the range of parameters in its potential do
not have a natural explanation.  Whether or not there are
fundamental scalar particles in nature is a question which will
hopefully be answered by experiments.  Superstring theories and
their low energy supergravities compactified to lower than 10
space-time dimensions normally contain a plethora of scalar fields
with a potential whose structure is tightly constrained by
supersymmetry.  The difficulty with such models is that they have
Anti-de-Sitter space as their most symmetric solution, that the
Calabi-Yau compactifications of 10 dimensional superstrings lack
in uniqueness, and that it is not clear what mechanism will prefer
and select such complicated manifolds from simpler flat 10
dimensional space-times.

The situation is different in  N=1 gauged supergravities (also
known as (1,0) type models). In D=6, such theories have the
minimum number of possible supersymmetries, namely 8 real
supercharges. By construction, they are chiral and therefore they
can give rise to interesting low energy chiral effective gauge
theories in  D=4. But perhaps more interestingly, they do not
admit the flat 6 dimensional space-time as a solution. In fact, it
has been shown recently \cite{Gibbons} that their unique maximally
symmetric ground state solution with a compact smooth
2-dimensional internal manifold is nothing but $R^{1,3}\times
S^2$. There are solutions of this type which also preserve half of
the D=6 supersymmetries  \cite{Salam} . The low energy D=4 theory
is thus a N=1 supergravity model.

These models being chiral are potentially inconsistent unless the
chiral anomaly cancels.  So far only one anomaly free model of
this type is known \cite{RD}. This model will be the framework for
our analysis and, to the extent that we need for our present
discussion, it will be explained in Section 2.

In this paper, we show that the scalar potential of the particular
class of models which we shall examine has the correct structure
to generate chaotic inflation. Inflation is triggered by complex
scalar fields coming from hyper-matter multiplets. The parameters
of the potential are fixed completely by supersymmetry. We shall
show that, provided a lower bound on the radius of
compactification is satisfied, we can obtain an acceptable mass
and acceptable self couplings of the scalar field which can give
rise to viable inflationary cosmology of chaotic inflation type
\cite{chaotic}.

One of the distinguishing features of the supergravity models that
we consider is that they have a global SU(2) R-symmetry. One can
gauge either the entire SU(2) or a U(1) subgroup of it. In the
following we shall consider a specific model in which a U(1)
subgroup of the SU(2) is gauged. We shall denote this U(1) as
$U(1)_R$.  Our specific choice is dictated by anomaly cancellation
in D=6. So far only one anomaly free model of this type has been
constructed and this is the model that we shall use as our
prototype example, although what we shall say will be applicable
to a much larger class of gauged (1,0) models in D=6.

%%%%%%%%%%%%%%%%%%%%%%%%%%%%%%%%%%%%%%%%%%%%%%%%%%%%%%%%%%%%%%%%%%%%

\section{The Model}

The specific model which we shall study belongs to the class of
gauged $N=1$ supergravity models. The Lagrangian for such models
have been constructed in \cite{Nishino1,Nishino2}. Being chiral,
they are potentially inconsistent due to the presence of gauge,
gravity and mixed anomalies. In \cite{RD}, an anomaly free model
with a gauge group $E_6\times E_7\times U(1)_R$ has been
constructed which involves a hyper-scalar multiplet transforming
in a 912-dimensional pseudo-real representation of $E_7$. So far,
this model remains the only known anomaly free example of a gauged
$N=1$ supergravity in D=6. The multiplet structure and a summary
of the known solutions of this model together with some of its
details have been given in a recent paper \cite{RD2} and will not
be repeated here (see also \cite{Cliff}).

For our present purpose we need only to know the following facts:
i) The bosonic sector consists of gravity, an antisymmetric tensor
potential of rank two, a dilaton field $\sigma$ and a  multiplet
of  912 complex scalar fields $\phi$ belonging to an irreducible
representation of $E_7$ , plus of course  the vector potentials in
the adjoint of the gauge group.

ii) In a convenient parameterization, the potential for the
scalars take a very simple form
\begin{equation} \label{sixpot}
V= \frac{g_1^2}{\kappa^4}
e^{-\kappa\sigma}[(1+|\phi|^2)^2+\frac{g_7 ^2}{g_1 ^2 }(\phi
T^a\phi)^2]
\end{equation}
where $\kappa$ is the D=6 gravitational coupling and has a
dimension of square length. The $g_1$ and $g_7$ are the coupling
constants of the $U(1)_R$ and $ E_7$, respectively, and they have
the dimensions of a length. Finally, the $T^a$ are the Hermitian
generators of $E_7$ in its 912-dimensional representation. The
most important property of this potential is that it has a unique
minimum at $\phi=0$. Hence, there is no moduli problem in this
model.

iii) Although the D=6 supersymmetry does not allow a bare
cosmological constant in the action, the value of the potential at
its minimum acts as a cosmological constant. This important fact
permits us to obtain a very simple compactifying solution for
which the extra two dimensions cover a $S^2$ and the remaining
four dimensions cover the flat Minkowski space time $R^{1,3}$. To
obtain this solution one or more of the U(1) gauge potentials
should assume a magnetic monopole configuration on $S^2$. It has
recently been shown in \cite{Gibbons} that these maximally
symmetric solutions are unique. The fact that the model itself
forces $R^{3,1}\times S^2$ as a ground state solution rather than
$R^{5,1}$ or an Anti-de-Sitter space is one of the most attractive
features of this class of models. Its uniqueness makes it even
more interesting. This should be compared with the non-uniqueness
of Calabi-Yau compactifications of string theories in D=10.

iv) If the monopole configuration resides in the $U(1)_R$
component of the gauge group the background preserves half of the
D=6 supersymmetries and  thus the effective D=4 theory will be an
anomaly free $N=1$ chiral supergravity. All other monopole
embeddings break supersymmetry completely.

%%%%%%%%%%%%%%%%%%%%%%%%%%%%%%%%%%%%%%%%%%%%%%%%%%%%%%%%%%%%%%%%%%%%%%
 \section{Inflation}
%%%%%%%%%%%%%%%%%%%%%%%%%%%%%%%%%%%%%%%%%%%%%%%%%%%

Within the context of effective field theories originating from
superstring or M theory, it has so far been difficult to obtain
realistic models with a transient phase of  cosmological
acceleration (see e.g. \cite{Susskind}), and even successful
models typically have a number of e-foldings of inflation which is
insufficient \cite{KKLMMT}.  Our D=6 supergravity theory contains
a lot of scalar fields with a potential which in principle should
produce inflation. Since the parameters entering this potential
are highly constrained by supersymmetry, we would like to see if
there is an allowed range in the space of these parameters such
that the effective 4-dimensional theory produces a viable
inflating universe.

From the 4-dimensional point of view, the bosonic sector of the
model will include gravitational modes as well  as scalar fields
coming from $\phi$ and its Kaluza-Klein (KK) modes. The KK modes
will belong to various representations of the SU(2) isometry group
of $S^2$  .  The details of the scalar potential will depend on
which  combination of $U(1)$ subgroups we identify with the
magnetic monopole configuration.  If the monopole resides in
$U(1)_R \times E_7$, then the Laplacian acting on $\phi$ will have
no SU(2) singlet zero mode (unless the magnetic charges coming
from the $U(1)_R$ and the $E_7$ factors cancel out precisely)
\cite{Randjbar-Daemi:1982hi} and the potential will mix all the
non-singlet KK modes.  On the other hand, if the monopole sits in
the $E_6$ factor of the gauge group, since the $\phi's$ are
singlets with respect to this group, the Laplacian will have a
zero mode which can be treated independently from the higher KK
modes by simply setting these modes to zero.  The potential for
the zero mode will simply be given by the same expression  as
(\ref{sixpot}), except that the constant term coming from 1 on the
right hand side will be absent, i.e.
\begin{equation} \label{fourpot}
V \, = \, 4\pi a^2 \frac{g_1^2}{\kappa^4}
e^{-\kappa\sigma}[(2|\phi|^2 + |\phi|^4+\frac{g_7 ^2}{g_1 ^2
}(\phi T^a\phi)^2] \, ,
\end{equation}
where $a$ is the radius of the $S^2$ and the factor of $4\pi a^2$
originates from integrating over $S^2$. In this potential,
$\sigma$ denotes the zero mode of the dilaton with respect to the
Laplacian, and it is a $SU(2)$ singlet.

In the case that $\phi$  couples to a net magnetic monopole of
charge $n$, its lowest mode  will be in a (2n + 1)-dimensional
representation of $SU(2)$   \cite{Randjbar-Daemi:1982hi} and
therefore it should carry appropriate $SU(2)$ indices which we
suppress.

Here we shall consider only the lowest lying modes of the
Laplacian acting on the $\phi's$ and will set to zero all the
higher modes.  The potential for such modes will have the general
structure given above.  It is not hard to work out its details and
write the expressions exactly as it has been done for some
examples in \cite{Dvali:2001qr} . We shall not need such details
for our discussion so we will restrict our attention to the
formula (\ref{fourpot}).

The background solution establishes a relationship between
$\kappa, g_1$, the constant vacuum expectation value $\sigma$ of
the dilaton, and the radius $a$ of $S^2$,  which is
\begin{equation} \label{radius}
a^2 \, = \, 2\frac{\kappa^2}{g_1^2} e^{\kappa\sigma} \, .
\end{equation}
Note also that given $\kappa$, the radius $a$ and the
four-dimensional Planck mass $m_{pl}$ are related via
\begin{equation} \label{Planck}
m_{pl}^2 \, = \, 4 \pi a^2 \kappa^{-2} \, .
\end{equation}
Using the relation (\ref{radius}), and also redefining $\phi$ so
that it has a correct mass dimensions of 1 in D=4 (we replace
$\phi$ in (\ref{fourpot}) by $\phi / m_{pl}$) one can show that
the masses $m_{\phi}$ of the  lowest lying $\phi$ modes will be
(modulo a constant of the order one)
\begin{equation} \label{mass}
m_{\phi}^2 = \frac{1}{a^2} \, .
\end{equation}
The scalar potential for these modes  involves two quartic self
couplings which are (making use of (\ref{Planck}), and again
modulo a constant of the order one)
\begin{equation} \label{quartic}
\lambda_1 \, = \, \frac{ \kappa^2}{a^4} \,\, {\rm and} \,\,
\lambda_2 \, = \, \frac{g_7 ^2}{g_1^2}\frac{\kappa ^ 2} {a^4} \, .
\end{equation}

The potential (\ref{fourpot}) allows for chaotic inflation. For a
single scalar field, inflation takes place while the value of the
rescaled field $\phi$ satisfies $|\phi| > m_{pl}$ (it is for such
field values that the slow-roll approximation for the scalar field
evolution is self-consistent, see e.g. \cite{RHBrev} for a recent
review). As first realized in \cite{assisted} and recently
discussed in more detail in \cite{Ho}, the slow-rolling
approximation is self-consistent for an even larger range of field
values if the inflaton is a multiplet of $N \gg 1$ equivalent
scalar fields, as in our case. To see this, let us for simplicity
drop the quartic coupling terms, and take the absolute values of
all $N$ scalar fields to be comparable. In this case, it follows
from the Klein-Gordon equation for the i'th scalar field $\phi_i$
\begin{equation} \label{KG}
{\ddot \phi_i} + 3 H {\dot \phi_i} \, = \, - m_{\phi}^2 \phi_i \,
,
\end{equation}
coupled with the Friedman equation relating the Hubble expansion
rate $H$ with the potential energy, that the velocity ${\dot
\phi_i}$ is suppressed by a factor of $N^{-1/2}$ compared to the
result for single field inflation, and that hence the dominance of
the energy density by potential energy, one of the two slow-roll
conditions, remains valid for field values $|\phi|$ a factor of
$N^{1/2}$ smaller than for single field inflation. In our case,
since $N = 1824$, this is an important effect. In particular, it
implies that the stage of inflation relevant for forming the
observed large-scale structure of the Universe corresponds to
field values smaller than the Planck scale, and thus corrections
to the potential from non-renormalizable terms in the supergravity
action are negligible (see the discussion in the following
section). Given the values of the masses and coupling constants of
(\ref{mass}) and (\ref{quartic}), (and assuming that $g_1$ and
$g_7$ are of the same order of magnitude) it also implies that the
mass and quartic terms have comparable magnitudes for field values
relevant to the final 50 e-foldings of inflation.

In order for quantum vacuum fluctuations produced during the
period of inflation not to produce too large an amplitude of the
spectrum of cosmic microwave anisotropies \cite{COBE}, one must
have (see e.g. \cite{RHBrev})
\begin{equation} \label{cond1}
({{m_{\phi}} \over{m_{pl}}})^2 \, \leq \, 10^{-12}
\end{equation}
or
\begin{equation} \label{cond2}
\lambda_i \, \leq \, 10^{-12} \, ,
\end{equation}
depending on which term in the potential dominates the scalar
field dynamics at the time when scales of cosmological interest
today exit the Hubble radius during inflation. In our case, both
conditions reduce to the same fairly mild condition on the radius
of the internal $S^2$ manifold, namely $a m_{pl} \geq 10^6$ or
\begin{equation} \label{constraint}
{a \over {\kappa^{1/2}}} \, \geq \, 10^3 \, .
\end{equation}
Thus, the requirement is that the radius of $S^2$ is larger than
the 6-D gravitational length $\kappa^{1/2}$ by a factor of $10^3$.

%%%%%%%%%%%%%%%%%%%%%%%%%%%%%%%%%%%%%%%%%%%%%%%%%%%%%%%%%%%%%%%%%%%%%%%%%%%%%

\section{Discussion}

In this note we have studied one aspect of early Universe
cosmology in the only known anomaly-free, gauged (1,0)
supergravity in six dimensions , a model in which the unique
vacuum state is Minkowski space-time cross an internal $S^2$. We
have shown that inflationary dynamics of chaotic type
(``large-field'' inflation) consistent with the cosmological
constraints coming from the amplitude of cosmic microwave
anisotropies can be realized provided that the radius of the
internal space $S^2$ satisfies the constraint (\ref{constraint}).
Since the inflaton is a multiplet of $N = 1824$ scalar fields,
slow-roll inflation is consistent for much smaller field values
than would be the case for a single inflaton field, namely for
field values satisfying $\phi < N^{-1/2} m_{pl}$. It is important
to check that the potential used is consistent for such field
values. The breakdown of the validity of the renormalizable
potential at field values relevant for inflation is often a
problem for models of chaotic inflation. A necessary condition for
the applicability of the potential (\ref{fourpot}) is that the
potential energy density during inflation is smaller than the
four-dimensional Planck density. This condition is trivially
satisfied given the constraints (\ref{cond1}) and (\ref{cond2}).
Since our model comes from a higher dimensional theory, a more
reasonable (and more restrictive) condition is to demand that the
potential energy density during inflation is less than the
six-dimensional Planck energy density $\kappa^{-2}$. For field
values $\phi \sim m_{pl}$ this condition would only marginally be
satisfied. However, the fact that $N \gg 1$ alleviates this
potential problem, since for field values $\phi \sim N^{-1/2}
m_{pl}$ we have

\begin{equation}
V(\phi) \, \sim \, \kappa^{-2} \, .
\end{equation}

To obtain this result, we have assumed that all $N$ fields have
comparable magnitudes, and that hence $|\phi|^2$ is given by $N
|\phi_i|^2$ and $|\phi|^4$ is given by $N^2 |\phi_i|^4$, where
$\phi_i$ is the characteristic value of a single field.

However, there may be a more stringent constraint: even field
values larger than the typical gravitational energy scale could be
problematic, and since the gravitational scale of relevance is the
six-dimensional one, our approximations would not be under control
for field values needed for inflation. However, as emphasized
recently in \cite{Linderev}, in the case of a theory in which
there is a shift symmetry of the Lagrangian in the absence of the
potential energy term, there is no physical reason which demands
the field values to be limited to sub-Planckian values, and the
only constraints to be imposed are the above energetic ones. In
our model, apart from the local gauge symmetries the Lagrangian is
also invariant under the quaternionic group $Sp(456)$, which acts
non trivially in the hypermatter sector and which is  a much
larger group than the shift symmetry group. The arguments of
\cite{Linderev} should extend to theories with this larger
symmetry group. Thus, we have shown that the low energy field
equations which we have used are consistent for the field values
which give us inflation.

In this paper we examined the possibility of chaotic inflation in
the dimensionally reduced model. It will be interesting to see if
the 6-dimensional field equations generate inflation of the type
which the effective 4-dimensional theory can give rise to. In
particular, it will be interesting to see if one can find an
inflationary solution in which the radius of the extra two
dimensions remain stable while our 3-dimensional world inflates.
It has been shown a long time ago that  the quantum free energy of a
real scalar field can act as a source of a radiation dominated
universe of this kind \cite{Randjbar-Daemi:1983jz}, see also the
recent paper \cite{Bringmann:2003pz}. Whether an inflationary
solution of this kind exists is not known yet.

Note that this anomaly-free $D=6$ supergravity model admits an
easy realization of the gravitational baryogenesis mechanism
recently suggested in \cite{Stephon}. This mechanism makes use of
the gravitational anomaly \cite{Witten} to relate the divergence
of the baryon current to $\tilde R  R$, where $R$ stands for the
Riemann tensor. This source vanishes for a homogeneous and
isotropic metric, but it is non-vanishing in the presence of a
spectrum of gravitational waves, provided there is a source of
cosmological bi-refringence. This birefringence is in our model
generated by certain components of a second rank antisymmetric
tensor potential which is necessarily present in models of the
type we are considering. Details will be presented in a subsequent
publication. In conclusion, we have shown that the gauged
supergravity model of \cite{RD} quite naturally leads to a period
of chaotic inflation in the early Universe. Further cosmological
aspects of this model will be analyzed in a followup publication.

\bigskip
\noindent {\bf Acknowledgements}:

RB wishes to thank the organizers of the 2004 ICTP Spring School
on Strings for the invitation to lecture, and for their
hospitality during the visit in the course of which this project
was initiated. RB is supported by the Perimeter Institute for
Theoretical Physics, and in part (at Brown) by the US Department
of Energy under Contract DE-FG02-91ER40688, TASK~A. He thanks the
Perimeter Institute for providing a stimulating atmosphere which
facilitated the completion of this project.

%%%%%%%%%%%%%%%%%%%%%%%%%%%%%%%%%%%%%%%%%%%%%%%%%%%%%%%%%%%%%%%%%%%%%%%%%%%

\end{document}